\documentclass[a4paper]{article}

\usepackage{latexsym}
\usepackage{a4wide}
\usepackage{graphicx, subfigure} 
\usepackage{cite} 
\usepackage{tabularx}
\usepackage{array}
\usepackage{graphics}     
\usepackage{amsmath}
\usepackage{amssymb}
\usepackage{longtable} 
\usepackage{verbatim} 
\usepackage[table]{xcolor}
\usepackage{booktabs}
\usepackage{hyperref} 
\usepackage[utf8]{inputenc}
\usepackage{multirow}
\usepackage{soul}

\definecolor{light-gray}{gray}{0.90}

\begin{document}

\hfill {\tt CERN-TH-2019-048, IPM/P.A-544, MITP/19-026}  

\def\thefootnote{\fnsymbol{footnote}}
 
\begin{center}

\vspace{3.cm}

{\Large\bf {Update on the $b \to s$ anomalies\footnote{This is an addendum to Ref.~\cite{Hurth:2017hxg}, 
"Lepton Nonuniversality in Exclusive $b \to s \ell^+ \ell^-$ Decays'', and Ref.~\cite{Arbey:2018ics},    
"Hadronic and New Physics Contributions to $b \to s$ Transitions''.}}}

\setlength{\textwidth}{11cm}
                    
\vspace{2.cm}
{\large\bf  
A.~Arbey$^{\,a,}$\footnote{Also Institut Universitaire de France, 103 boulevard Saint-Michel, 75005 Paris, France}$^{,}$\footnote{Email: alexandre.arbey@ens-lyon.fr}, 
T.~Hurth$^{b,}$\footnote{Email: tobias.hurth@cern.ch},
F.~Mahmoudi$^{a,\dagger,}$\footnote{Email: nazila@cern.ch},\\
D.~Mart\'inez Santos$^{c,}$\footnote{Email: Diego.Martinez.Santos@cern.ch},
S.~Neshatpour$^{d,}$\footnote{Email: neshatpour@ipm.ir }
}
 
\vspace{1.cm}
{\em $^a$Univ Lyon, Univ Lyon 1, CNRS/IN2P3, Institut de Physique Nucl\'eaire de Lyon, UMR5822, F-69622 Villeurbanne, France}\\[0.2cm]
{\em $^b$PRISMA Cluster of Excellence and  Institute for Physics (THEP)\\
Johannes Gutenberg University, D-55099 Mainz, Germany}\\[0.2cm]
{\em $^c$Instituto Galego de F\'isica de Altas Enerx\'ias, Universidade de Santiago de Compotela, Spain}\\[0.2cm]
{\em $^d$School of Particles and Accelerators,
Institute for Research in Fundamental Sciences (IPM)
P.O. Box 19395-5531, Tehran, Iran}

\end{center}

\renewcommand{\thefootnote}{\arabic{footnote}}
\setcounter{footnote}{0}

\vspace{1.cm}
\thispagestyle{empty}
\centerline{\bf ABSTRACT}
\vspace{0.5cm}
{ We present a brief update of our model-independent analyses of the $b \to s$ data  presented in the articles published in Phys.~Rev.~D96~(2017)~095034 and  Phys.~Rev.~D98~(2018)~095027  based on new data on $R_K$ by  LHCb,  on $R_{K^*}$
by Belle, and  on $B_{s,d} \to \mu^+\mu^-$ by ATLAS.}

\newpage

\begin{figure}[!th]
\begin{center}
\includegraphics[width=0.45\textwidth]{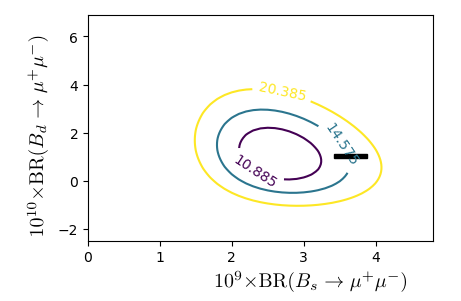}
\caption{\small 
2D likelihood plot where the contours are 1, 2 and $3\sigma$ (in terms of $\Delta\chi^2$).
The numbers correspond to the absolute $\chi^2$ and the black box is the SM prediction.
\label{fig:2Dlikelihood}}
\end{center}
\end{figure}

{\bf New data:}  Using the theoretical framework introduced in Refs.~\cite{Hurth:2017hxg,Arbey:2018ics}  we update our results in view of the following new experimental measurements: 
\begin{itemize}
\item The most awaited one is  the LHCb measurement of the lepton-universality testing observable $R_K\equiv {\rm BR}(B^+ \to K^+ \mu^+ \mu^-)/{\rm BR}(B^+ \to K^+ e^+ e^-)$.
The LHCb measurement using 5~fb$^{-1}$ of data~\cite{Aaij:2019wad} collected with the center of mass energies of 7,  8 and 13 TeV  for $R_K$ in the low-dilepton mass ($q^2$) bin leads to 
\begin{align}
 R_K([1.1,6.0]\,{\rm GeV}^2)= 0.846^{+0.060+0.016}_{-0.054-0.014}\;,
\end{align}
where the first and second uncertainties are the systematic and statistical errors, respectively. 
Compared to the previous LHCb measurement based on  3 fb$^{-1}$ of data~\cite{Aaij:2014ora}, the central value is now closer to the SM prediction,  but the significance of the tension is still $2.5\sigma$ due to the smaller uncertainty of the new measurement.
\item  Moreover, there has been new experimental results on another lepton-universality testing observable 
$R_{K^*}\equiv {\rm BR}(B \to K^* \mu^+ \mu^-)/{\rm BR}(B \to K^* e^+ e^-)$ by the Belle collaboration~\cite{Abdesselam:2019wac},
both for the neutral and charged $B$ mesons. The results are given in three low-$q^2$ bins and one high-$q^2$ bin which for the combined
charged and neutral channels are
\begin{align}
 R_{K^*}([0.045,1.1]\,{\rm GeV}^2)= 0.52^{+0.36}_{-0.26}\pm 0.05,\quad 
 R_{K^*}([1.1,6.0]\,{\rm GeV}^2)&= 0.96^{+0.45}_{-0.29} \pm 0.11, \nonumber\\  
 R_{K^*}([0.1,8]\,{\rm GeV}^2)= 0.90^{+0.27}_{-0.21} \pm 0.10, \quad
 R_{K^*}([15,19]\,{\rm GeV}^2)&= 1.18^{+0.52}_{-0.32} \pm 0.10.
\end{align} 
For our analysis  we consider the $[0.1,8]$ GeV$^2$ bin (together with the high-$q^2$ bin) and do not use the very low $q^2$ bin below 0.1 GeV$^2$ as advocated by Ref.~\cite{Bordone:2016gaq} in order to avoid near-threshold uncertainties which would be present when the lower range of the bin is set to the di-muon threshold.

We note  that the Belle measurement for the low-$q^2$ bin, $[0.045,1.0]$, which we do not use, has a tension with the SM prediction which is slightly more than $1\sigma$, while the other bins are
all well in agreement with the SM at the $1\sigma$-level. All the $R_{K^*}$ measurements of Belle  are in agreement with the LHCb measurement~\cite{Aaij:2017vbb} due to the large uncertainties of the Belle results.
\item  Our update also takes into account new experimental data on  $B_{s,d} \to \mu^+ \mu^-$ by ATLAS~\cite{Aaboud:2018mst}.   
We have combined this new result with the previous results of CMS~\cite{CMS:2014xfa} and LHCb~\cite{Aaij:2017vad} 
building a joint 2D likelihood (see Fig.~\ref{fig:2Dlikelihood}) with common $f_d/f_s$ and ${\rm BR}(B^+\to J/\,\psi K^+) \times {\rm BR}(J/\psi\to \mu^+ \mu^-)$
which finally leads us to 
\begin{align}
 {\rm BR}(B_s \to \mu^+ \mu^-) = 2.65_{-0.39}^{+0.43} \times 10^{-9}, \quad \quad \quad  {\rm BR}(B_d \to \mu^+ \mu^-) = 1.09_{-0.68}^{+0.74} \times 10^{-10}.
\end{align}

\end{itemize}

The calculation of the observables is performed with SuperIso v4.1 \cite{Mahmoudi:2007vz}. The statistical methods used for our study are described in \cite{Hurth:2014vma,Hurth:2016fbr}. In particular, we compute the theoretical covariance matrix for all the observables and consider the experimental correlations provided by the experiments. For the hadronic corrections, we do not consider hadronic parameters as in Refs.~\cite{Chobanova:2017ghn,Arbey:2018ics} but use 10\% error assumption as explained in~\cite{Hurth:2016fbr}.

{\bf Comparison of $R_K$ and $R_{K^*}$ data with other $b \to s$ data:} 
The hadronic contributions which are usually the main source of theoretical uncertainty cancel out in the case of the
potentially lepton flavour violating ratios $R_K$ and $R_{K^*}$ and thus, very precise predictions are possible in the SM~\cite{Hiller:2003js}. In contrast, the power corrections to  the angular observables and other observables in the exclusive $b \to s$ sector are still  not really under control  and are usually guesstimated to $10\%$, $20\%$ or even higher percentages of the leading nonfactorisable contributions to those observables. However, there is a promising approach based on analyticity, which may lead to a clear estimate of such effects and which may  allow for a clear separation of hadronic and new physics (NP) effects in these observables~\cite{Bobeth:2017vxj}.

As argued in Ref.~\cite{Hurth:2017hxg}, the present situation suggests separate analyses of the theoretically very clean ratios and the other $b \to s$ observables.  In Table~\ref{tab:ComparisonRKRKstar_1D}, the one-operator fits to new physics have been compared  when considering all the relevant data on $b\to s$ transitions except for $R_K$ and $R_{K^*}$ and when only considering the data on $R_K$ and $R_{K^*}$\footnote{
The right (left) hand side results of Table~\ref{tab:ComparisonRKRKstar_1D} in this paper give the updated results of Table 1 (2) in Ref.~\cite{Hurth:2017hxg}
where here we have not normalised to the SM values.}. 
We note that the NP significance of the ratios is reduced compared to our previous  analysis~\cite{Hurth:2017hxg}, mainly because of the new measurements  of $R_{K^*}$ by  Belle which are compatible with the SM predictions at the $1\sigma$-level as stated above.  But within the one-operator fits we find again that the NP analyses of the two sets of observables are less coherent than
often stated, especially regarding the coefficients $C_{10}^{\mu,e}$.

\begin{table}[th!]
\begin{center}
\setlength\extrarowheight{0pt}
\hspace*{-1.cm}
\scalebox{0.80}{
\hspace*{.9cm}
\begin{tabular}{|l|r|r|c|}
\hline 
\multicolumn{4}{|c|}{\footnotesize All observables except $R_K, R_{K^*}$ \vspace{-0.1cm}} 	\\
\multicolumn{4}{|c|}{\footnotesize ($\chi^2_{\rm SM}=100.2$)}\\ \hline
                          & b.f. value & $\chi^2_{\rm min}$ & ${\rm Pull}_{\rm SM}$  \\ 
\hline \hline
$\delta C_{9} $          		& $ 	-1.00	\pm	0.20	 $ & $	82.5	 $ & $ 	4.2	\sigma	 $  \\
$\delta C_{9}^{\mu} $      		& $ 	-1.03	\pm	0.20	 $ & $	80.3	 $ & $ 	4.5	\sigma	 $  \\
$\delta C_{9}^{e} $        		& $ 	0.72	\pm	0.58	 $ & $	98.9	 $ & $ 	1.1	\sigma	 $  \\
\hline
$\delta C_{10} $         		& $ 	0.25	\pm	0.23	 $ & $	98.9	 $ & $ 	1.1	\sigma	 $  \\
$\delta C_{10}^{\mu} $     		& $ 	0.32	\pm	0.22	 $ & $	98.0	 $ & $ 	1.5	\sigma	 $  \\
$\delta C_{10}^{e} $       		& $ 	-0.56	\pm	0.50	 $ & $	99.1	 $ & $ 	1.0	\sigma	 $  \\
\hline
$\delta C_{\rm LL}^\mu$			& $ 	-0.48	\pm	0.15	 $ & $	89.1	 $ & $ 	3.3	\sigma	 $  \\
$\delta C_{\rm LL}^e$			& $ 	0.33	\pm	0.29	 $ & $	99.0	 $ & $ 	1.1	\sigma	 $  \\
\hline
\end{tabular}
\hspace*{2.cm}
\begin{tabular}{|l|r|r|c|}
\hline 
\multicolumn{4}{|c|}{\small Only $R_K, R_{K^*}$ \vspace{-0.1cm}} 	\\
\multicolumn{4}{|c|}{\small ($\chi^2_{\rm SM}=16.9$)}\\ \hline
                          & b.f. value & $\chi^2_{\rm min}$ & ${\rm Pull}_{\rm SM}$  \\ 
\hline \hline
$\delta C_{9} $          		& $ 	-2.04	\pm	5.93	 $ & $	16.8	 $ & $ 	0.3	\sigma	 $  \\
$\delta C_{9}^{\mu} $      		& $ 	-0.74	\pm	0.28	 $ & $	8.4	 $ & $ 	2.9	\sigma	 $  \\
$\delta C_{9}^{e} $        		& $ 	0.79	\pm	0.29	 $ & $	7.7	 $ & $ 	3.0	\sigma	 $  \\
\hline
$\delta C_{10} $         		& $ 	4.10	\pm	11.87	 $ & $	16.7	 $ & $ 	0.5	\sigma	 $  \\
$\delta C_{10}^{\mu} $     		& $ 	0.77	\pm	0.26	 $ & $	6.1	 $ & $ 	3.3	\sigma	 $  \\
$\delta C_{10}^{e} $       		& $ 	-0.78	\pm	0.27	 $ & $	6.0	 $ & $ 	3.3	\sigma	 $  \\
\hline
$\delta C_{\rm LL}^\mu$			& $ 	-0.37	\pm	0.12	 $ & $	7.0	 $ & $ 	3.1	\sigma	 $  \\
$\delta C_{\rm LL}^e$ 			& $ 	0.41	\pm	0.15	 $ & $	6.8	 $ & $ 	3.2	\sigma	 $  \\
\hline
\end{tabular}
} 
\caption{\small Comparison of one-operator NP fits where the $\delta C_{\rm LL}^\ell$ basis corresponds to $\delta C_{9}^{\ell}=-\delta C_{10}^{\ell}$.
On the left hand side all relevant data on $b \to s$ transitions except $R_K$ and $R_{K^*}$ (with 10\% error assumption for the power corrections) is used and
on the right hand side only the data on $R_K, R_{K^*}$ is  considered.
\label{tab:ComparisonRKRKstar_1D}} 
\end{center} 
\end{table}
\begin{table}[th!]
\begin{center}
\setlength\extrarowheight{0pt}
\hspace*{-1.cm}
\scalebox{0.80}{
\hspace*{.9cm}
\begin{tabular}{|l|r|r|c|}
\hline 
 \multicolumn{4}{|c|}{\footnotesize All observables except $R_K, R_{K^*}, B_{s,d}\to \mu^+ \mu^-$ \vspace{-0.1cm}} \\ 
 \multicolumn{4}{|c|}{\footnotesize ($\chi^2_{\rm SM}=99.7$)} \\ \hline
                          & b.f. value & $\chi^2_{\rm min}$ & ${\rm Pull}_{\rm SM}$  \\ 
\hline \hline
$\delta C_{9} $          	& $ 	-1.03	\pm	0.20	 $ & $	81.0	 $ & $ 	4.3	\sigma	 $  \\
$\delta C_{9}^{\mu} $      	& $ 	-1.05	\pm	0.19	 $ & $	78.8	 $ & $ 	4.6	\sigma	 $  \\
$\delta C_{9}^{e} $        	& $ 	0.72	\pm	0.58	 $ & $	98.5	 $ & $ 	1.1	\sigma	 $  \\
\hline
$\delta C_{10} $         	& $ 	0.27	\pm	0.28	 $ & $	98.7	 $ & $ 	1.0	\sigma	 $  \\
$\delta C_{10}^{\mu} $     	& $ 	0.38	\pm	0.28	 $ & $	97.7	 $ & $ 	1.4	\sigma	 $  \\
$\delta C_{10}^{e} $       	& $ 	-0.56	\pm	0.50	 $ & $	98.7	 $ & $ 	1.0	\sigma	 $  \\
\hline
$\delta C_{\rm LL}^\mu$ 	& $ 	-0.50	\pm	0.16	 $ & $	88.8	 $ & $ 	3.3	\sigma	 $  \\
$\delta C_{\rm LL}^e$ 		& $ 	0.33	\pm	0.29	 $ & $	98.6	 $ & $ 	1.1	\sigma	 $  \\
\hline
\end{tabular}
\hspace*{2.cm}
\begin{tabular}{|l|r|r|c|}
\hline 
 \multicolumn{4}{|c|}{\footnotesize Only $R_K, R_{K^*}, B_{s,d}\to \mu^+ \mu^-$ \vspace{-0.1cm}} \\
 \multicolumn{4}{|c|}{\footnotesize ($\chi^2_{\rm SM}=19.0$)} \\ \hline
                          & b.f. value & $\chi^2_{\rm min}$ & ${\rm Pull}_{\rm SM}$  \\ 
\hline \hline
$\delta C_{9} $          		& $ 	-2.04	\pm	5.93	 $ & $	18.9	 $ & $ 	0.3	\sigma	 $  \\
$\delta C_{9}^{\mu} $      		& $ 	-0.74	\pm	0.28	 $ & $	10.6	 $ & $ 	2.9	\sigma	 $  \\
$\delta C_{9}^{e} $        		& $ 	0.79	\pm	0.29	 $ & $	9.9	 $ & $ 	3.0	\sigma	 $  \\
\hline
$\delta C_{10} $         		& $ 	0.43	\pm	0.32	 $ & $	17.0	 $ & $ 	1.4	\sigma	 $  \\
$\delta C_{10}^{\mu} $     		& $ 	0.65	\pm	0.20	 $ & $	6.9	 $ & $ 	3.5	\sigma	 $  \\
$\delta C_{10}^{e} $       		& $ 	-0.78	\pm	0.27	 $ & $	8.2	 $ & $ 	3.3	\sigma	 $  \\
\hline
$\delta C_{\rm LL}^\mu$ 		& $ 	-0.37	\pm	0.11	 $ & $	7.2	 $ & $ 	3.4	\sigma	 $  \\
$\delta C_{\rm LL}^e$			& $ 	0.41	\pm	0.15	 $ & $	9.0	 $ & $ 	3.2	\sigma	 $  \\
\hline
\end{tabular}
} 
\caption{\small Comparison of one operator NP fits where the $\delta C_{\rm LL}^\ell$ basis corresponds to $\delta C_{9}^{\ell}=-\delta C_{10}^{\ell}$.
On the left hand side all relevant data on $b \to s$ transitions except $R_K, R_{K^*}, B_{s,d}\to \mu^+ \mu^-$ 
(with 10\% error assumption for the power corrections) is used and
on the right hand side only the data on $R_K, R_{K^*}, B_{s,d}\to \mu^+ \mu^-$ is  considered. 
\label{tab:ComparisonRKRKstarBmumu_1D}} 
\end{center} 
\end{table}
%
%

One may expect that the observables $B_{s,d} \to \mu^+\mu^-$ are responsible  for the finding that NP in $C_{10}^{\mu,e}$ is favoured in the fit to the ratios $R_{K^{(*)}}$ but not in the fit to the rest of the $b\to s$ transitions. However, when besides $R_K, R_{K^*}$ also the $B_{s,d}\to \mu^+ \mu^-$ observables are removed from the rest of the $b \to s$ observables and compared to the fit when considering the data on $R_K, R_{K^*},B_{s,d}\to \mu^+ \mu^-$ we find that at least within the one-operator fits the observables $B_{s,d} \to \mu^+\mu^-$ do not play a major role: The results in Table~\ref{tab:ComparisonRKRKstarBmumu_1D} are very similar with the ones 
in Table~ \ref{tab:ComparisonRKRKstar_1D}. This feature is consistent with our finding in Ref.~\cite{Hurth:2017hxg} that 
the observables $B_{s,d} \to \mu^+\mu^-$ will not play a primary role in the future differentiation between the NP hypotheses for the ratios $R_{K^{(*)}}$.  However, with the new average for BR($B_s \to \mu^+\mu^-$) which includes the ATLAS measurement, there is a  tension of $1.5\sigma$ with the SM prediction which suggests the same direction for $C_{10}^\mu$ as it is preferred by the $R_{K^{(*)}}$ fit.
This can also be seen by comparing the right hand sides of Tables~\ref{tab:ComparisonRKRKstar_1D} and~\ref{tab:ComparisonRKRKstarBmumu_1D} where there is a slight increase in the SM-Pull when the data on 
$B_s \to \mu^+\mu^-$ is added to the  $R_{K^{(*)}}$ fit.

\begin{figure}[!t]
\begin{center}
\includegraphics[width=0.33\textwidth]{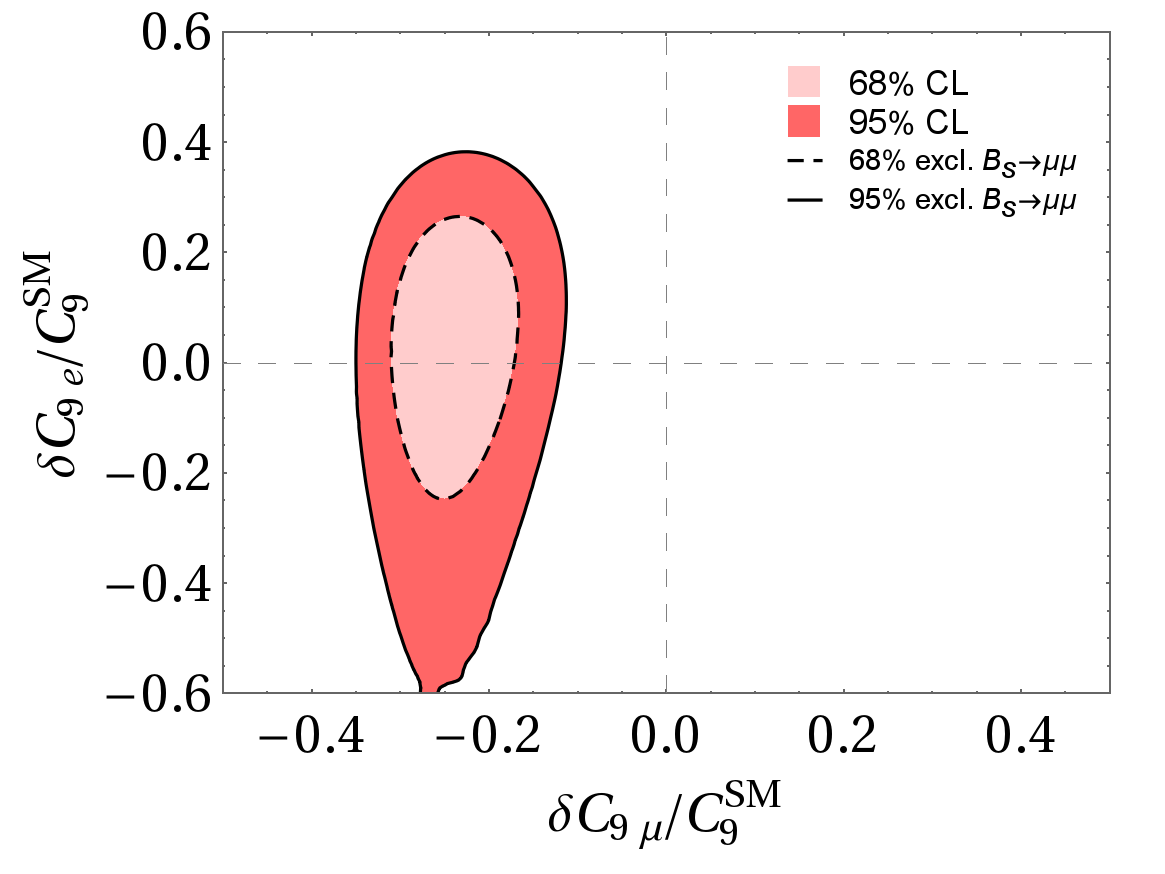}
\includegraphics[width=0.32\textwidth]{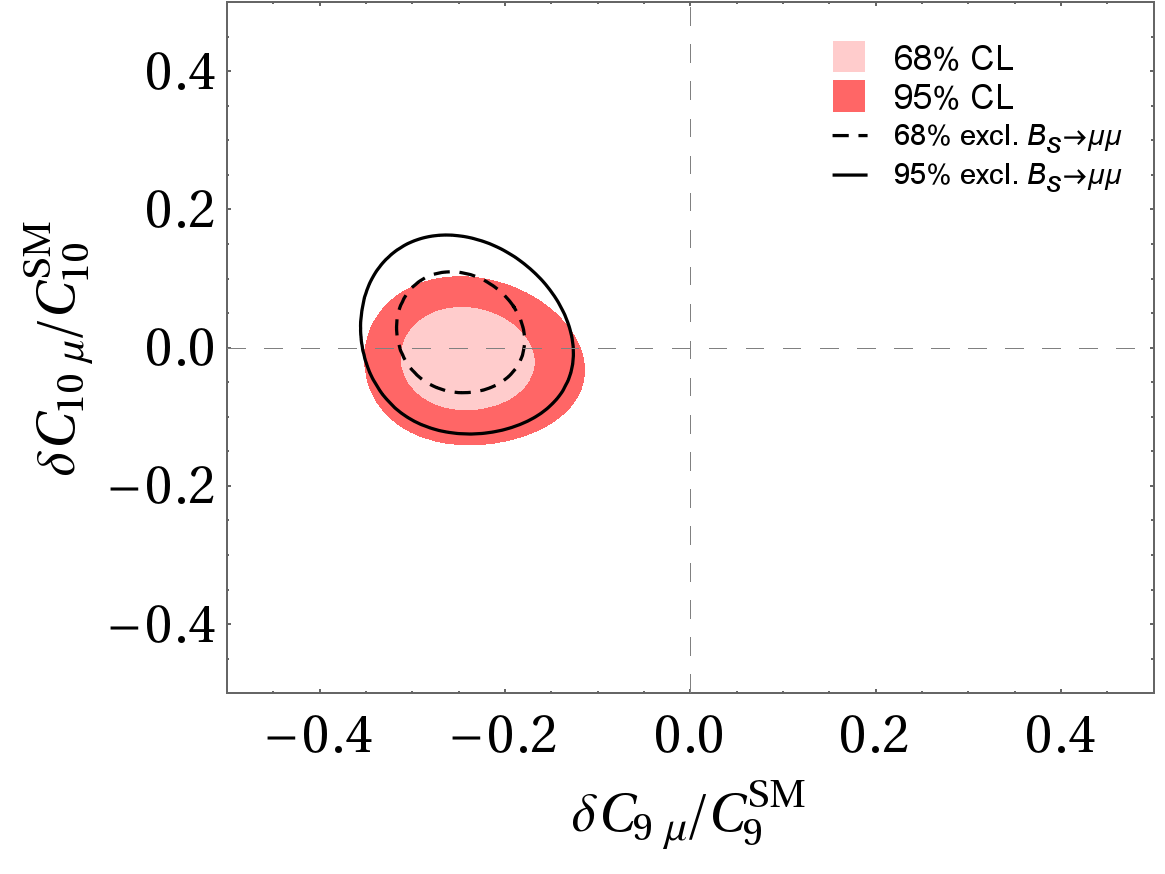}
\includegraphics[width=0.32\textwidth]{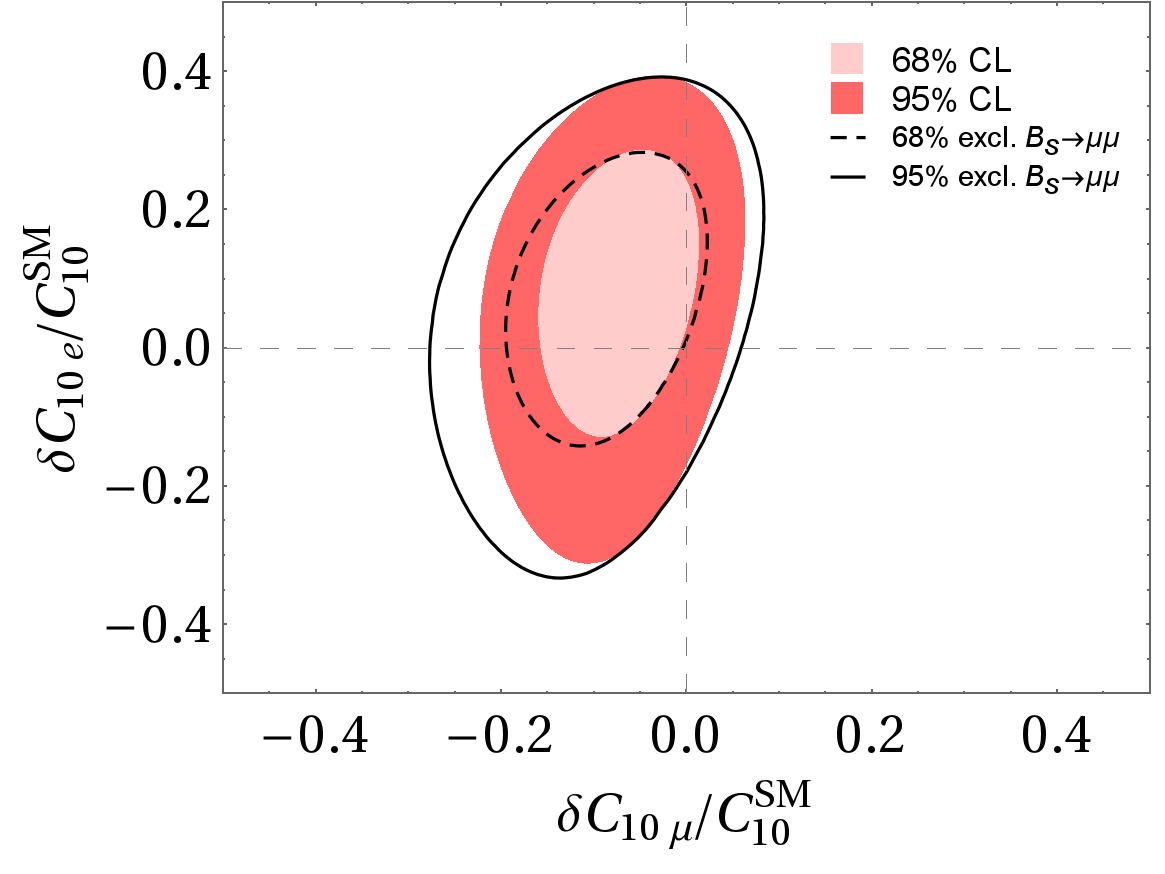}
\\[2.mm]
\includegraphics[width=0.33\textwidth]{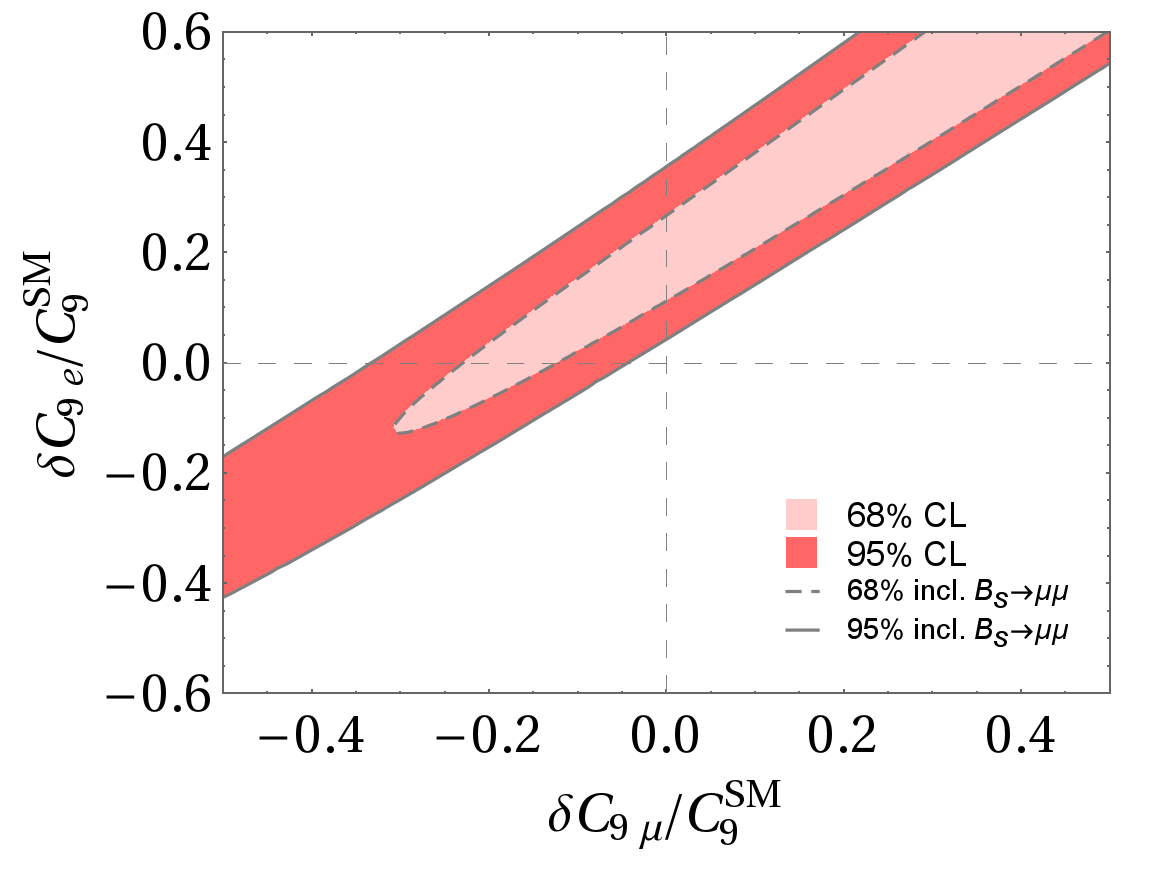}
\includegraphics[width=0.32\textwidth]{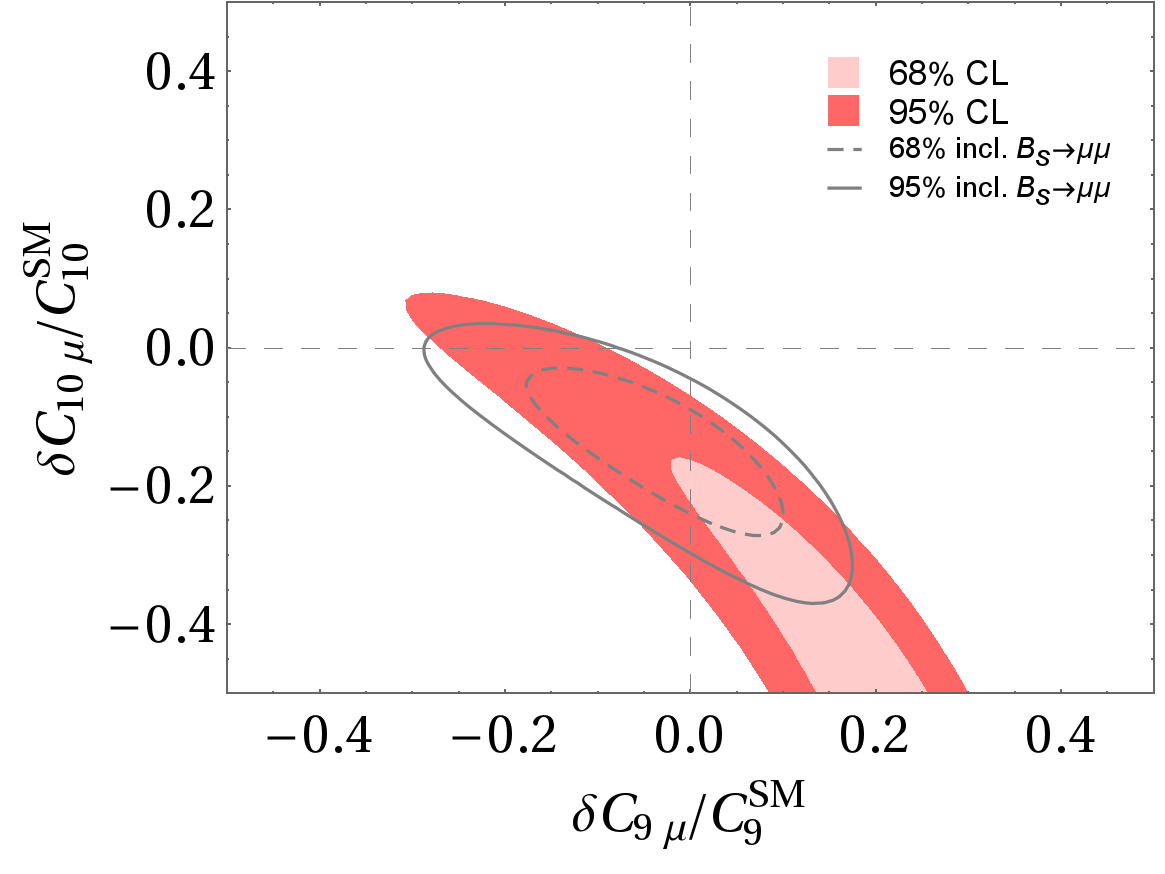}
\includegraphics[width=0.32\textwidth]{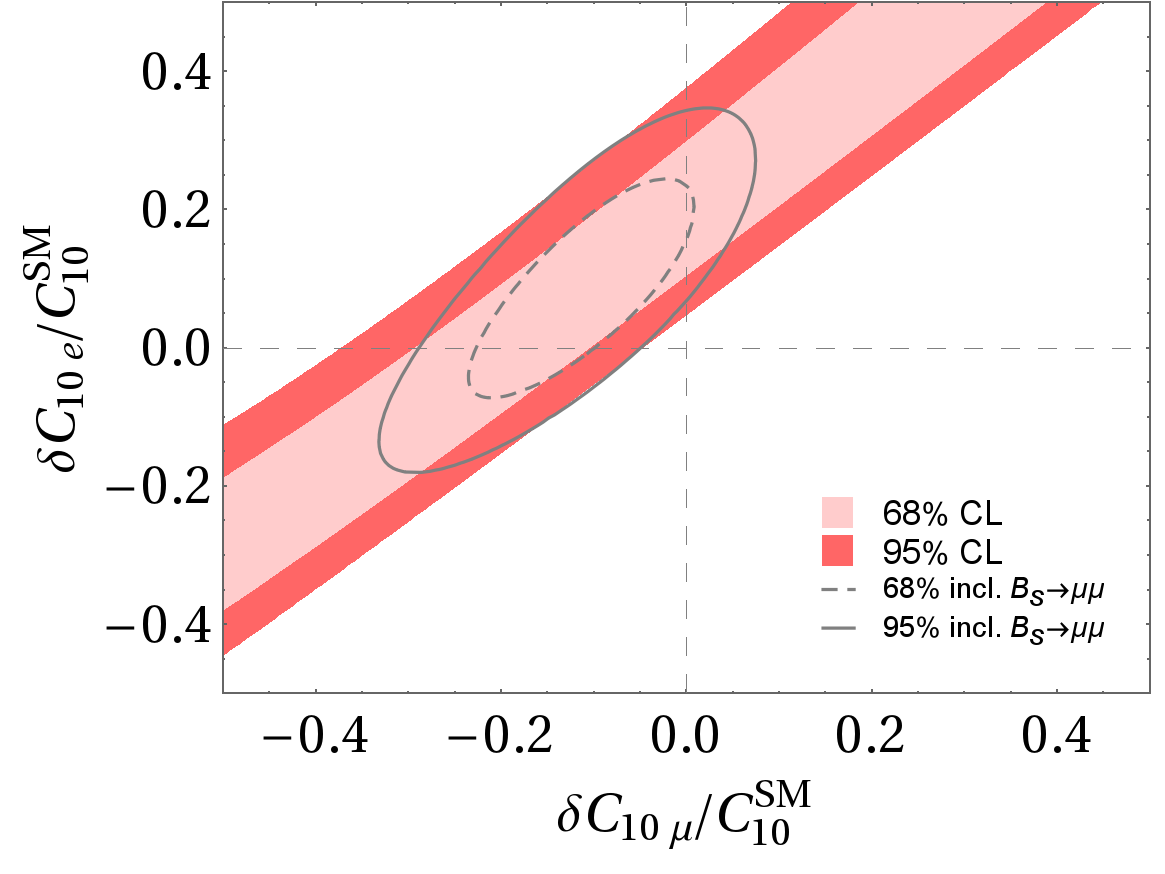}
\caption{\small Two operator fits to NP. The contours correspond to the 68 and 95\% confidence level regions. 
On the upper row we have considered all observables except $R_K$ and $R_{K^*}$ with the assumption of 10\% power corrections.
On the lower row we have only used the data on $R_K, R_{K^*}$.
Pull$_{\rm SM}$ for the 1$^{\rm st}$, 2$^{\rm nd}$, 3$^{\rm rd}$ column are respectively, $4.1, 4.1, 1.1\sigma$ ($3.1, 3.2, 3.1\sigma$), 
for the upper (lower) plots.
The black (gray) dashed and solid contours correspond to excluding (including) the data on $B_{s,d}\to \mu^+ \mu^-$ from (to) the fits of the upper (lower) plots.
\label{fig:RKRKstarBmumu}}
\end{center}
\end{figure}

In the next step we compare the two sets of observables in two-operator fits.  In Fig.~\ref{fig:RKRKstarBmumu} the two operator fits for $\{C_{9}^e, C_9^\mu \}$, $\{C_{10}^\mu, C_9^\mu \}$, and for $\{C_{10}^\mu, C_{10}^e \}$ are shown, using only the data on $R_K, R_{K^*}$, 
or all observables except $R_K, R_{K^*}$  where the effect of moving the data on $B_{s,d}\to \mu^+ \mu^-$ observables from one set to the other has been shown with the black and gray contours. The latter ones nicely show the influence of these observables when more than one operator is considered. Independent of these effects one finds that the two sets of observables are compatible at least at the $2\sigma$-level. 

%
\begin{table}[th!]
\begin{center}
\setlength\extrarowheight{0pt}
\hspace*{-1.cm} 
\scalebox{0.80}{
\begin{tabular}{|l|r|r|c|}
\hline 
 \multicolumn{4}{|c|}{\small All observables  ($\chi^2_{\rm SM}=117.03$)} \\ \hline
                          & b.f. value & $\chi^2_{\rm min}$ & ${\rm Pull}_{\rm SM}$  \\ 
\hline \hline
$\delta C_{9} $          	& $ 	-1.01	\pm	0.20	 $ & $ 	99.2	 $ & $	4.2	\sigma	 $  \\
$\delta C_{9}^{\mu} $      	& $ 	-0.93	\pm	0.17	 $ & $ 	89.4	 $ & $	5.3	\sigma	 $  \\
$\delta C_{9}^{e} $        	& $ 	0.78	\pm	0.26	 $ & $ 	106.6	 $ & $	3.2	\sigma	 $  \\
\hline
$\delta C_{10} $         	& $ 	0.25	\pm	0.23	 $ & $ 	115.7	 $ & $	1.1	\sigma	 $  \\
$\delta C_{10}^{\mu} $     	& $ 	0.53	\pm	0.17	 $ & $ 	105.8	 $ & $	3.3	\sigma	 $  \\
$\delta C_{10}^{e} $       	& $ 	-0.73	\pm	0.23	 $ & $ 	105.2	 $ & $	3.4	\sigma	 $  \\
\hline
$\delta C_{\rm LL}^\mu$ 	& $ 	-0.41	\pm	0.10	 $ & $ 	96.6	 $ & $	4.5	\sigma	 $  \\
$\delta C_{\rm LL}^e$		& $ 	0.40	\pm	0.13	 $ & $ 	105.8	 $ & $	3.3	\sigma	 $  \\
\hline
\end{tabular}
} 
\caption{\small Best fit values and errors in the one operator fits to all the relevant data on $b \to s$ transitions, assuming 10\% error for the power corrections. 
\label{tab:ALL_1D}} 
\end{center} 
\end{table}

\begin{figure}[!ht]
\begin{center}
\includegraphics[width=0.33\textwidth]{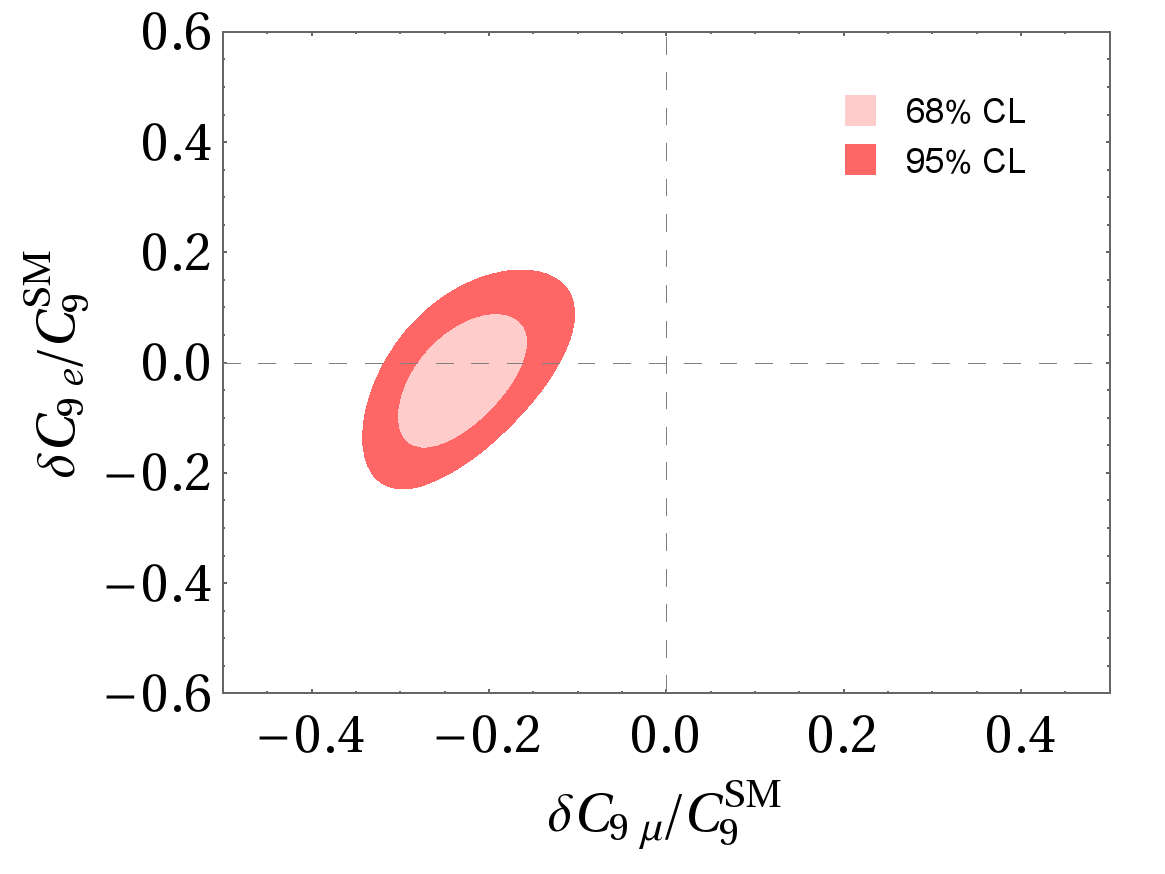}
\includegraphics[width=0.32\textwidth]{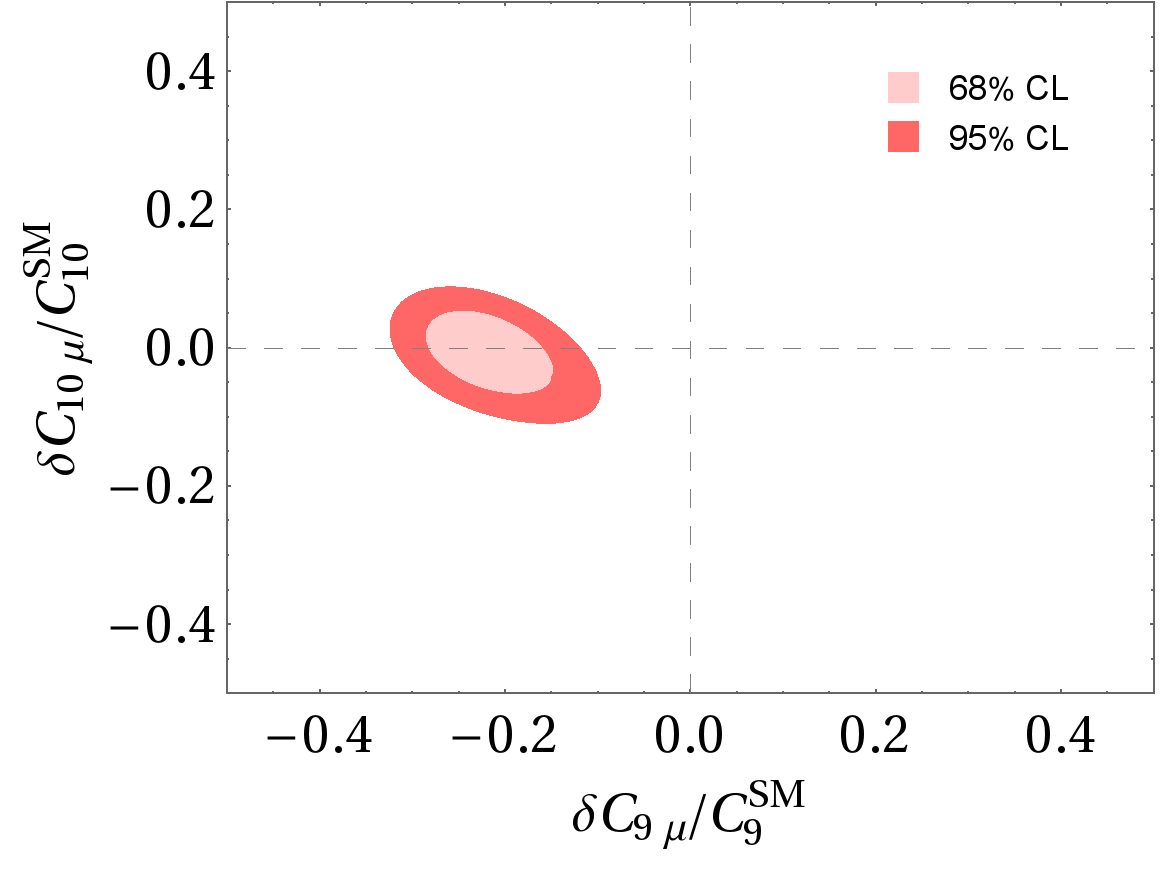}
\includegraphics[width=0.32\textwidth]{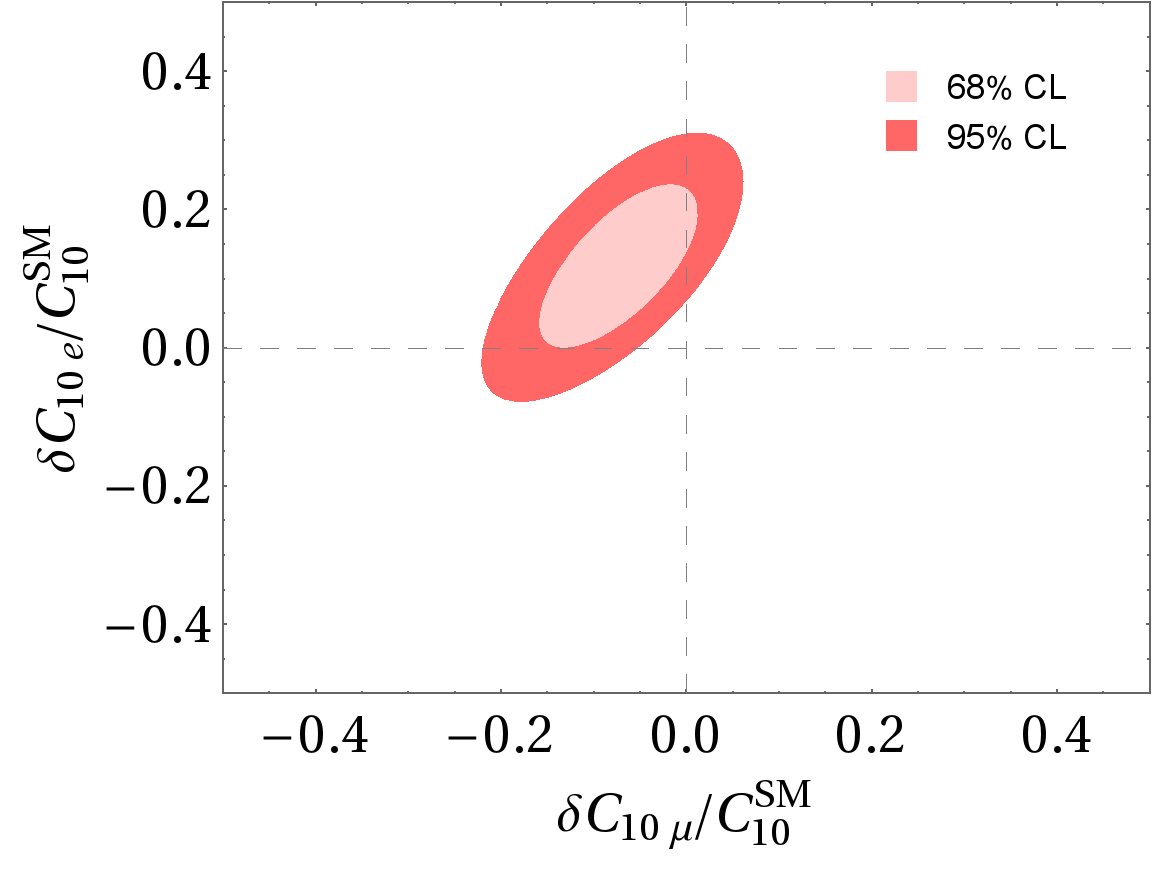}
\\[2.mm]
\includegraphics[width=0.32\textwidth]{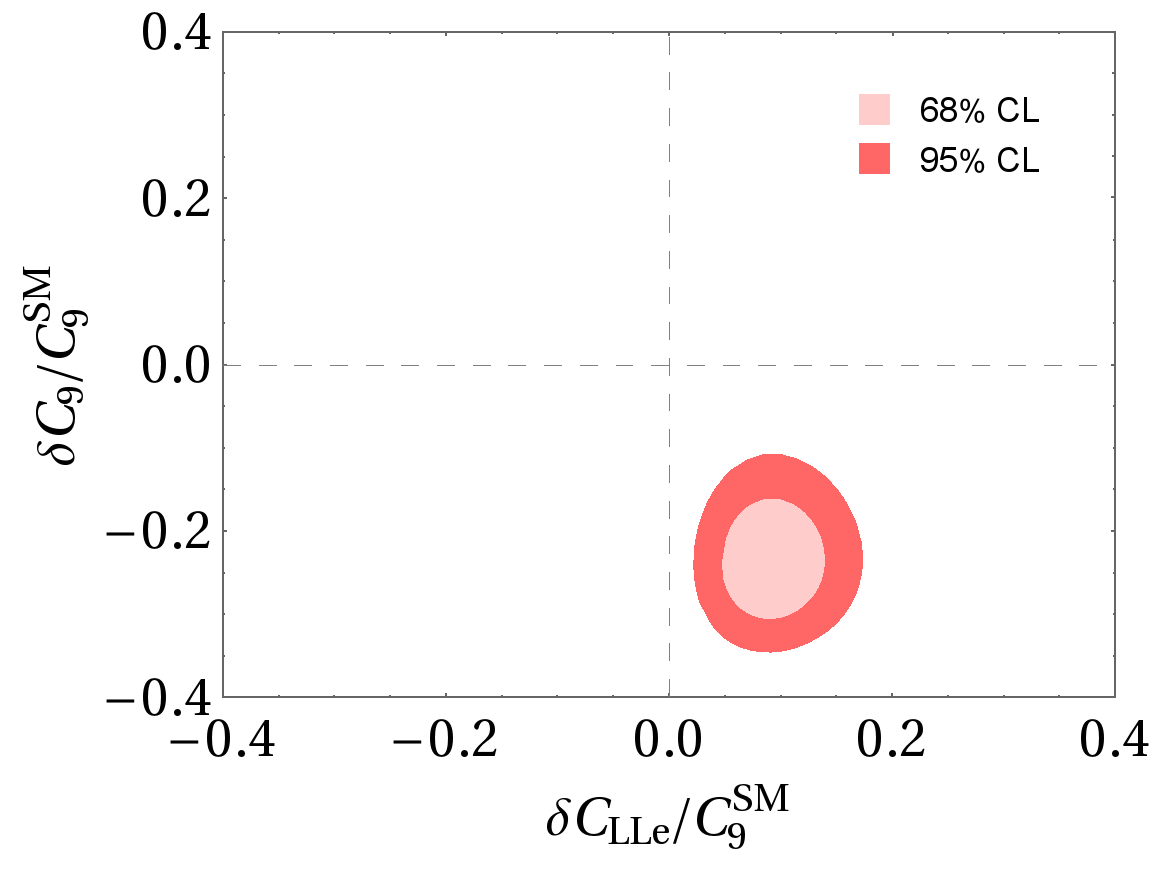}\qquad\includegraphics[width=0.32\textwidth]{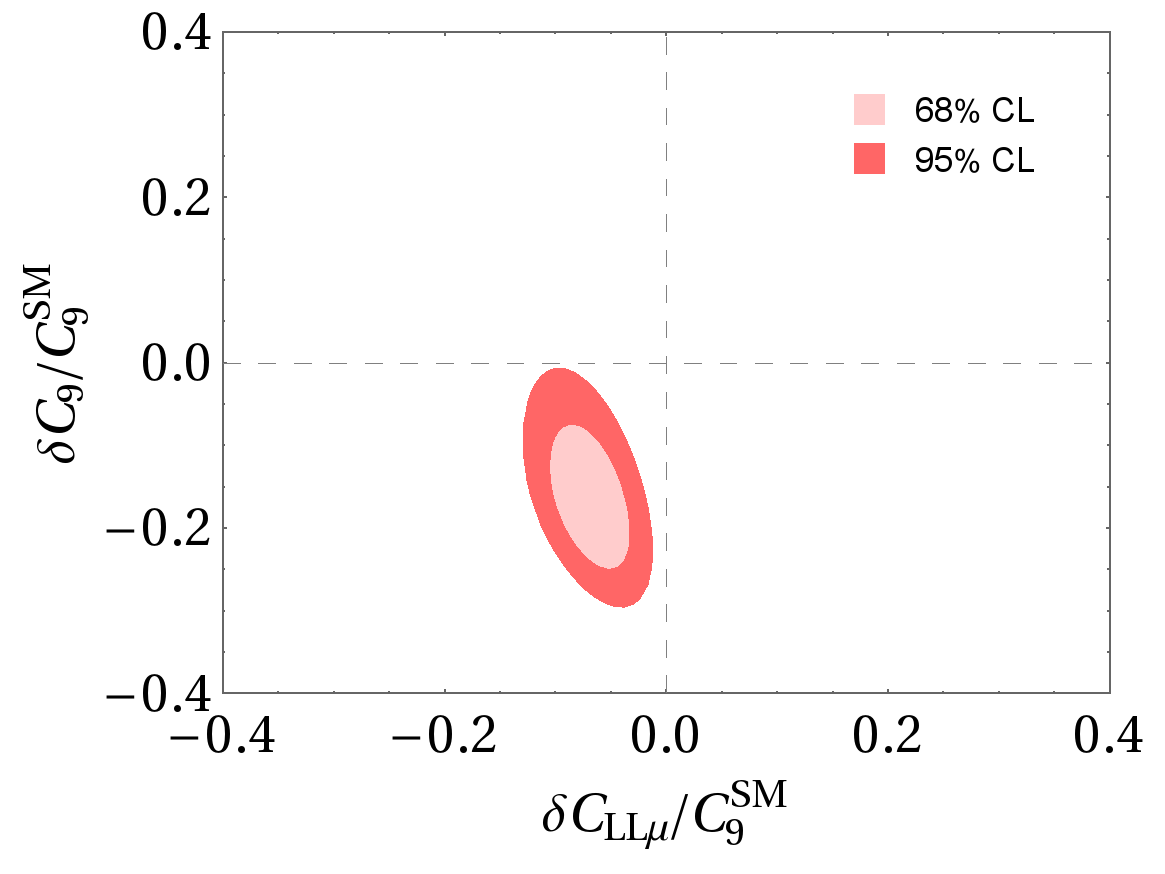}
\caption{\small Two operator fits to NP, considering all observables (with the assumption of 10\% power corrections). 
Pull$_{\rm SM}$ in the $\{C_{9}^e , C_{9}^\mu\}, \{C_{10}^\mu , C_{9}^\mu\}, \{C_{10}^{e} , C_{10}^{\mu}\}$ fits are $4.9, 4.9, 3.2\sigma$, respectively.
Pull$_{\rm SM}$ for the $\{C_{LL}^{e},C_9\}$ and $\{C_{LL}^{\mu},C_9\}$ fits of the lower row are $5.0$ and $4.8\sigma$, respectively.
\label{fig:all}}
\end{center}
\end{figure}
{\bf   Global fit}  In Table~\ref{tab:ALL_1D}, the global one-operator fits to NP are given where \emph{all} the relevant data on $b \to s$ transitions are considered\footnote{This table includes updated results of Table 5 in Ref.~\cite{Arbey:2018ics}.}.
In Fig.~\ref{fig:all}, the two operator fits for $\{C_{9}^e , C_{9}^\mu\}, \{C_{10}^\mu , C_{9}^\mu\}$ and $\{C_{10}^\mu , C_{10}^e\}$ (the same set as in Fig.~\ref{fig:RKRKstarBmumu}) can be seen. Moreover, the fits for $\{C_{LL}^{\mu},C_9\}$ and $\{C_{LL}^e,C_9\}$ are given which are also motivated for model building  (\emph{e.g.} see Ref.~\cite{Cornella:2019hct}). These fits are always done under the assumption of $10\%$ power corrections in the angular observables. Compared with our previous analysis in Ref.~\cite{Arbey:2018ics} the NP significance in the one- and also in the two-operator fits is reduced  by at least $0.5\sigma$. Only in cases of flavour-symmetric $C_9$ and $C_{10}$ which are independent from the changes in the ratios one finds the same NP significance as expected. 

\begin{table}[b!]
\begin{center}
\setlength\extrarowheight{3pt}
\scalebox{0.83}{
\begin{tabular}{|c|c|c|c|}
\hline 
  \multicolumn{4}{|c|}{All observables  with $\chi^2_{\rm SM}=117.03$} \\ 
  \multicolumn{4}{|c|}{$\chi^2_{\rm min}=71.96;\; {\rm Pull}_{\rm SM}=3.3\sigma (3.8\sigma)$} \\ 
\hline \hline
$\delta C_7$ & $\delta C_7^\prime$ & $\delta C_8$ & $\delta C_8^\prime$\\ 
$ -0.01 \pm 0.04 $ & $ 0.01 \pm 0.03 $ & $ 0.82 \pm 0.72 $ & $ -1.65 \pm 0.47 $\\ 
\hline\hline  
$\delta C_{9}^{\mu}$ & $\delta C_{9}^{e}$ & $\delta C_{10}^{\mu}$ & $\delta C_{10}^{e}$ \\
$ -1.37 \pm 0.25 $ & $ -6.55 \pm 2.37 $ & $ -0.11 \pm 0.27 $ & $ 2.34 \pm 3.11 $  \\
\hline 
$\delta C_{9}^{\prime \mu}$ & $\delta C_{9}^{\prime e}$ & $\delta C_{10}^{\prime \mu}$ & $\delta C_{10}^{\prime e}$ \\
$ 0.23 \pm 0.62 $ & $ 0.75 \pm 2.82 $ & $ -0.16 \pm 0.36 $ & $ 1.67 \pm 3.05 $ \\
\hline \hline
$C_{Q_{1}}^{\mu}$ & $C_{Q_{1}}^{e}$ & $C_{Q_{2}}^{\mu}$ & $C_{Q_{2}}^{e}$ \\ 
$ -0.01 \pm 0.09 $ & undetermined & $ -0.05 \pm 0.19 $ & undetermined  \\
\hline 
$C_{Q_{1}}^{\prime \mu}$ & $C_{Q_{1}}^{\prime e}$ & $C_{Q_{2}}^{\prime \mu}$ & $C_{Q_{2}}^{\prime e}$ \\ 
$ 0.13 \pm 0.09 $ & undetermined & $ -0.18 \pm 0.20 $ &  undetermined   \\
\hline
\end{tabular}
} 
\caption{Best fit values for the 20 operator global fit to the $b \to s$ data, assuming 10\% error for the power corrections.
Pull$_{\rm SM} = 3.3\sigma (3.8\sigma)$ when considering 20 (16) degrees of freedom. The number in the parenthesis corresponds to the effective number of degrees of freedom in which the insensitive coefficients are not counted (see Ref.~\cite{Arbey:2018ics} for more details). 
\label{tab:ALL_20D_C78910C12primes}} 
\end{center} 
\end{table}
%

The observables $B_{s,d} \to \mu^+ \mu^-$ are usually used to strongly constrain NP effects in scalar and pseudoscalar operators. As a consequence, a general usage is to consider the contributions from the scalar and pseudoscalar as vanishingly small. However, as mentioned in Ref.~\cite{Arbey:2018ics}, this is only valid when the relation between the scalar and pseudoscalar operators ($C_{Q_{1}}=-C_{Q_{2}}$) is assumed, which breaks the possible degeneracy between $C_{Q_2}$ and $C_{10}$ and allows for strong constraints on $C_{Q_{1,2}}$. In general scenarios, $C_{Q_{2}}$ and $C_{10}$ can have simultaneously large values which compensates, while indeed $C_{Q_{1}}$ is rather constrained (for more details see Ref.~\cite{Arbey:2018ics}). Since beyond simplified NP models, there can be scenarios which contain various new particles and several new couplings we also perform a multidimensional fit in Table~\ref{tab:ALL_20D_C78910C12primes} where all the relevant Wilson coefficients which amounts to 20 coefficients are modified\footnote{This table updates the results given in Table 8 of Ref.~\cite{Arbey:2018ics}.}.

Finally, we note that there have been  other model-independent  analyses presented recently which update previous analyses~\cite{Hurth:2017hxg,Arbey:2018ics,Capdevila:2017bsm,DAmico:2017mtc,Altmannshofer:2017yso,Ciuchini:2017mik,Geng:2017svp} based on the new experimental data. We find small differences with these updated analyses~\cite{DAmico:2017mtc,Alguero:2019ptt,Alok:2019ufo,Ciuchini:2019usw,Aebischer:2019mlg,Kowalska:2019ley} only in the NP significances. This can be explained by the different choices of bins in the new Belle measurement and by slightly different treatments of power corrections and  of form factors. 

In summary, the overall picture of the $b \to s$ anomalies remains the same as before taking into account  the new results from LHCb, Belle and ATLAS on $R_K, R_{K^*}$ and  $B_s \to \mu^+\mu^-$.  Although, the significance of the new physics description of the $R_K{^{(*)}}$ data is now  reduced by more than half a $\sigma$.  Nevertheless, the future measurements of these theoretically very clean ratios and similar observables which are sensitive to lepton flavour non-universality have a great potential to unambiguously establish lepton non-universal new physics.


\providecommand{\href}[2]{#2}\begingroup\raggedright
\endgroup


\begin{thebibliography}{10}
\small{
\bibitem{Hurth:2017hxg}
  T.~Hurth, F.~Mahmoudi, D.~Martinez Santos and S.~Neshatpour,
  Phys.\ Rev.\ D {\bf 96} (2017) no.9,  095034
  [arXiv:1705.06274 [hep-ph]].
  
\bibitem{Arbey:2018ics}
  A.~Arbey, T.~Hurth, F.~Mahmoudi and S.~Neshatpour,
  Phys.\ Rev.\ D {\bf 98} (2018) no.9,  095027
  [arXiv:1806.02791 [hep-ph]].

\bibitem{Aaij:2019wad}
  R.~Aaij {\it et al.} [LHCb Collaboration],
  arXiv:1903.09252 [hep-ex].

\bibitem{Aaij:2014ora}
  R.~Aaij {\it et al.} [LHCb Collaboration],
  Phys.\ Rev.\ Lett.\  {\bf 113} (2014) 151601
  [arXiv:1406.6482 [hep-ex]].  
  
\bibitem{Abdesselam:2019wac}
  A.~Abdesselam {\it et al.} [Belle Collaboration],
  arXiv:1904.02440 [hep-ex].

\bibitem{Bordone:2016gaq}
  M.~Bordone, G.~Isidori and A.~Pattori,
  Eur.\ Phys.\ J.\ C {\bf 76} (2016) no.8,  440
  [arXiv:1605.07633 [hep-ph]].  
  
\bibitem{Aaij:2017vbb}
  R.~Aaij {\it et al.} [LHCb Collaboration],
  JHEP {\bf 1708} (2017) 055
  [arXiv:1705.05802 [hep-ex]].  
  
\bibitem{Aaboud:2018mst}
  M.~Aaboud {\it et al.} [ATLAS Collaboration],
  [arXiv:1812.03017 [hep-ex]].  

\bibitem{CMS:2014xfa}
  V.~Khachatryan {\it et al.} [CMS and LHCb Collaborations],
  Nature {\bf 522} (2015) 68
  [arXiv:1411.4413 [hep-ex]].    

\bibitem{Aaij:2017vad}
  R.~Aaij {\it et al.} [LHCb Collaboration],
  Phys.\ Rev.\ Lett.\  {\bf 118} (2017) no.19,  191801
  [arXiv:1703.05747 [hep-ex]].  

 

\bibitem{Mahmoudi:2007vz}
  F.~Mahmoudi,
  Comput.\ Phys.\ Commun.\  {\bf 178} (2008) 745
  [arXiv:0710.2067 [hep-ph]]; 
  Comput.\ Phys.\ Commun.\  {\bf 180} (2009) 1579
  [arXiv:0808.3144 [hep-ph]]; 
  Comput.\ Phys.\ Commun.\  {\bf 180} (2009) 1718.

\bibitem{Hurth:2014vma}
  T.~Hurth, F.~Mahmoudi and S.~Neshatpour,
  JHEP {\bf 1412} (2014) 053
  [arXiv:1410.4545 [hep-ph]].
  
\bibitem{Hurth:2016fbr}
  T.~Hurth, F.~Mahmoudi and S.~Neshatpour,
  Nucl.\ Phys.\ B {\bf 909} (2016) 737
  [arXiv:1603.00865 [hep-ph]].
  
\bibitem{Chobanova:2017ghn}
  V.~G.~Chobanova, T.~Hurth, F.~Mahmoudi, D.~Martinez Santos and S.~Neshatpour,
  JHEP {\bf 1707} (2017) 025
  [arXiv:1702.02234 [hep-ph]].
  
\bibitem{Hiller:2003js}
  G.~Hiller and F.~Kruger,
  Phys.\ Rev.\ D {\bf 69} (2004) 074020
  [hep-ph/0310219].

\bibitem{Bobeth:2017vxj}
  C.~Bobeth, M.~Chrzaszcz, D.~van Dyk and J.~Virto,
  Eur.\ Phys.\ J.\ C {\bf 78} (2018) no.6,  451
  [arXiv:1707.07305 [hep-ph]].

\bibitem{Cornella:2019hct}
  C.~Cornella, J.~Fuentes-Martin and G.~Isidori,
  arXiv:1903.11517 [hep-ph].
  
\bibitem{Capdevila:2017bsm}
  B.~Capdevila, A.~Crivellin, S.~Descotes-Genon, J.~Matias and J.~Virto,
  JHEP {\bf 1801} (2018) 093
  [arXiv:1704.05340 [hep-ph]].  

\bibitem{Altmannshofer:2017yso}
  W.~Altmannshofer, P.~Stangl and D.~M.~Straub,
  Phys.\ Rev.\ D {\bf 96} (2017) no.5,  055008
  [arXiv:1704.05435 [hep-ph]].
  
\bibitem{Ciuchini:2017mik}
  M.~Ciuchini, A.~M.~Coutinho, M.~Fedele, E.~Franco, A.~Paul, L.~Silvestrini and M.~Valli,
  Eur.\ Phys.\ J.\ C {\bf 77} (2017) no.10,  688
  [arXiv:1704.05447 [hep-ph]].  


\bibitem{Geng:2017svp}
  L.~S.~Geng, B.~Grinstein, S.~Jager, J.~Martin Camalich, X.~L.~Ren and R.~X.~Shi,
  Phys.\ Rev.\ D {\bf 96} (2017) no.9,  093006
  [arXiv:1704.05446 [hep-ph]].

\bibitem{DAmico:2017mtc}
  G.~D'Amico, M.~Nardecchia, P.~Panci, F.~Sannino, A.~Strumia, R.~Torre and A.~Urbano,
  JHEP {\bf 1709} (2017) 010
  [arXiv:1704.05438 [hep-ph]].    

\bibitem{Alguero:2019ptt}
  M.~Algueró, B.~Capdevila, A.~Crivellin, S.~Descotes-Genon, P.~Masjuan, J.~Matias and J.~Virto,
  arXiv:1903.09578 [hep-ph].  

\bibitem{Alok:2019ufo}
  A.~K.~Alok, A.~Dighe, S.~Gangal and D.~Kumar,
  arXiv:1903.09617 [hep-ph].
  
\bibitem{Ciuchini:2019usw}
  M.~Ciuchini, A.~M.~Coutinho, M.~Fedele, E.~Franco, A.~Paul, L.~Silvestrini and M.~Valli,
  arXiv:1903.09632 [hep-ph].

  \bibitem{Aebischer:2019mlg}
  J.~Aebischer, W.~Altmannshofer, D.~Guadagnoli, M.~Reboud, P.~Stangl and D.~M.~Straub,
  arXiv:1903.10434 [hep-ph].

\bibitem{Kowalska:2019ley}
  K.~Kowalska, D.~Kumar and E.~M.~Sessolo,
  arXiv:1903.10932 [hep-ph].

  }
\end{thebibliography}
\end{document}